\documentclass[aps,twocolumn,superscriptaddress]{revtex4-1}

\usepackage{amsmath,amssymb,color,graphicx,mathpazo,mathtools,times,float,dsfont,mathrsfs}
\usepackage[colorlinks=true,allcolors=blue]{hyperref}
\usepackage{soul}
\usepackage{comment}
\definecolor{Red}{rgb}{1,0,0}

\usepackage[textsize=scriptsize]{todonotes}

\begin{document}

\title{Photonic frequency conversion of OFDM microwave signals in a wavelength-scale optomechanical cavity}

%\maketitle

% Author: Please give full first and last names for authors and include * after the name of all corresponding authors

\author{Laura Mercad\'e}
\affiliation{Nanophotonics Technology Center, Universitat Polit\`ecnica de Val\`encia, Camino de Vera s/n, 46022 Valencia, Spain}
\author{Maria Morant}
\affiliation{Nanophotonics Technology Center, Universitat Polit\`ecnica de Val\`encia, Camino de Vera s/n, 46022 Valencia, Spain}
\author{Amadeu Griol}
\affiliation{Nanophotonics Technology Center, Universitat Polit\`ecnica de Val\`encia, Camino de Vera s/n, 46022 Valencia, Spain}
\author{Roberto Llorente}
\affiliation{Nanophotonics Technology Center, Universitat Polit\`ecnica de Val\`encia, Camino de Vera s/n, 46022 Valencia, Spain}
\author{Alejandro Mart\'inez}
\email{amartinez@ntc.upv.es}
\affiliation{Nanophotonics Technology Center, Universitat Polit\`ecnica de Val\`encia, Camino de Vera s/n, 46022 Valencia, Spain}

% Dedication

%\dedication{Optional dedication here. If no dedication is required, please leave blank}

% Affiliations: Please provide adacemic titles (Prof. or Dr.) for all authors where applicable, and include an institutional email address for all corresponding authors

\date{\today}% It is always \today, today,


% Keywords: Please provide a minimum of three and a maximum of seven keywords, separated by commas

%\keywords{Keyword 1, Keyword 2, Keyword 3}

% Abstract should be written in the present tense and impersonal style (i.e., avoid we), and be at most 200 words long
\begin{abstract}
%\justifying
 Optomechanical (OM) cavities enable coupling of near-infrared light and GHz-frequency acoustic waves in wavelength-scale volumes. When driven in the phonon lasing regime, an OM cavity can perform simultaneously as a nonlinear mixer and a local oscillator --at integer multiples of the mechanical resonance frequency-- in the optical domain. In this work, we use this property to demonstrate all-optical frequency down- and up-conversion of MHz-bandwidth orthogonal frequency division multiplexed signals compliant with the IEEE 802.16e WiMAX wireless standard at microwave frequencies. To this end, we employ a silicon OM crystal cavity (OMCC), supporting a breathing-like mechanical resonance at $f_m\approx$ 3.9 GHz and having a foot-print \hbox{$\approx$ 10 $\mu$m$^{2}$}, which yields frequency conversion efficiencies better than \hbox{-17 dB} in both down- and up-conversion processes at mW-scale driving power. This work paves the way towards the application of OMCCs in low-power all-photonic processing of digitally-modulated microwave signals in miniaturized silicon photonics chips.

\end{abstract}
\maketitle
% Text: Please use section headings and subheadings as specified below. For communications, all section headings apart from Experimental Section should be removed
% Please make the first reference to a display item bold: \textbf{Figure 1}
% Do not abbreviate Figure, Equation, etc.; display items are always singular, i.e., Figure 1 and 2.
% Equations are always singular, i.e., Equation 1 and 2, and should be inserted using the {equation} environment, not as graphics
% Please do not use footnotes in the text, additional information can be added to the Reference list.

\section{Introduction}
%\justifying
Cavity optomechanics studies how light and mechanical waves interact with each other when confined in a cavity \cite{KIP08-SCI,ASP14-RMP,PEN14-NP}. Such interaction leads to a wide variety of physical phenomena, many of them related to the possibility to either attenuate (cooling \cite{CHAN11-NAT}) or amplify (heating \cite{MA11-NAT}) the mechanical motion of the cavity. Indeed, under blue-detuned laser driving of the cavity, the motion can be amplified up to a point that mechanical losses are overcome and the cavity reaches a state of self-sustained oscillations \cite{ASP14-RMP, KIP05-PRL}. This closely resembles the lasing process for optical waves, so this phenomenon is usually called phonon lasing \cite{GRU10-PRL,NAV15-SR,VAL09-NAT}. Optomechanical cavities can be patterned in released high-index films using standard micro- and nano-fabrication tools (such as electron-beam lithography), thus coexisting on-chip with other photonics, electronics or mechanical components \cite{Li15-OPT,Jia19-OPT}. The resulting OM crystal cavities (OMCCs) \cite{EIC09-NAT} allow for the independent design of optical and mechanical resonances \cite{OUD14-PRB}, as well as to maximize its interaction strength \cite{MATH18-APL}, which is given by the OM coupling rate, $g_{0}$. Moreover, since the crystal periodicity imposes the wavelengths for which photons and phonons get coupled, telecom-band light can interact with GHz-scale mechanical motion, so OMCCs can play the role of ultra-compact elements linking optical and microwave signals \cite{VAI16-APL,MIR20-NAT}. This feature is particularly relevant in microwave photonics, the discipline that addresses the processing of radio-frequency (RF) signals at microwave frequencies in the optical domain \cite{CAP07-NPHOT,MAR19-NP}. This approach has multiple advantages compared to the processing in the electric domain, including nearly-unlimited bandwidth, immunity to electromagnetic interference and frequency-independent losses, amongst others, while displaying all the benefits of optical domain processing \cite{TAN20-LPR}.

\begin{figure*}[t]
\begin{center}
\includegraphics[width=1\linewidth]{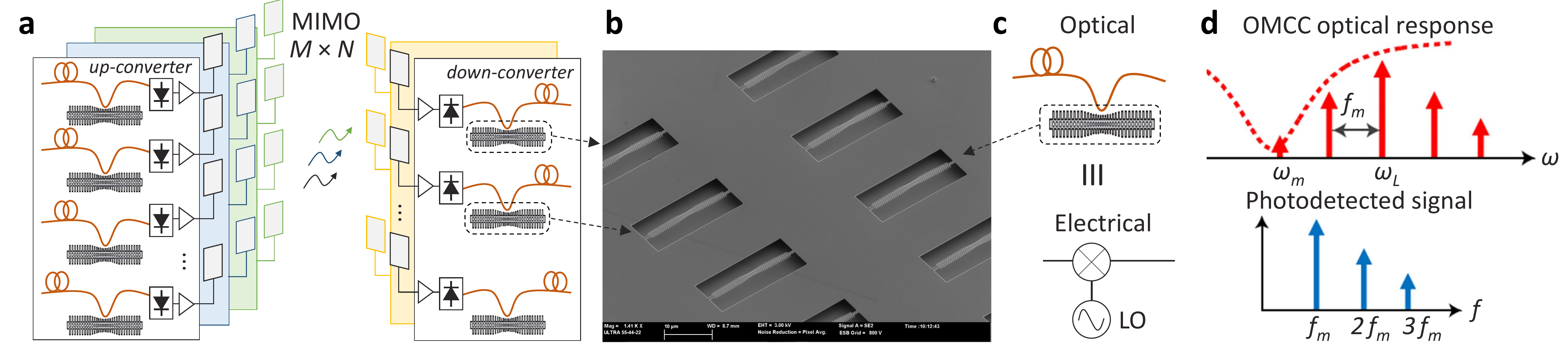}%
\end{center}
\caption{\textbf{a} Application scenario for up- and down-conversion of RF signals for wireless MIMO transmission with multiple AEs in a $N \times M$ arrangement. \textbf{b} Scanning Electron Microscopy (SEM) image image of an array of silicon OMCCs -- as the one used in this work -- which could be used for up/down-conversion in each AE. \textbf{c} Block diagram of the proposed all-optical up/down-converter and equivalent electrical implementation with an external mixer and an LO. \textbf{d} Optical response of the OMCC providing several harmonics at integer multiples of the mechanical resonance frequency and photodetected signal in the electrical domain.}
\label{fig:motivation}
\end{figure*}

A prominent example of microwave signal processing is the realization of frequency up- and down-conversion of data streams extensively performed in wireless networks \cite{TAN20-LPR}. In particular, such conversion processes are essential in multiple-input multiple-output (MIMO) radio transmission systems, as depicted in Fig. \ref{fig:motivation}a. Indeed, massive MIMO systems employ arrays with several antenna elements (AEs) at the transmitter and/or receiver (as depicted in the example shown in Fig. \ref{fig:motivation}a with $N \times M$ AEs). Nowadays, massive MIMO systems have been proposed as a key solution to meet the increasing demanding throughput of wireless and cellular networks \cite{RAT18}. Thus, it is foreseen that future wireless communications will rely on massive MIMO systems equipped with hundreds or even thousands of AEs. Although the fabrication of affordable large antenna arrays on a small footprint is already viable, the corresponding up/down-conversion chains --which include power-consuming mixers and local oscillators (LOs) as depicted in Fig. \ref{fig:motivation}c-- are bulky and expensive \cite{RAT18}. Therefore, solutions that comprehend extreme miniaturization, easy scalability, massive on-chip integration into arrays and low-power consumption are highly desirable \cite{MAR19-NP}. Interestingly, multiple oscillators built on the same chip may be synchronized in frequency by weak mechanical interaction whilst being coupled to different optical fields \cite{Colombano19}.

\begin{figure*}[t]
\begin{center}
\includegraphics[width=0.75\linewidth]{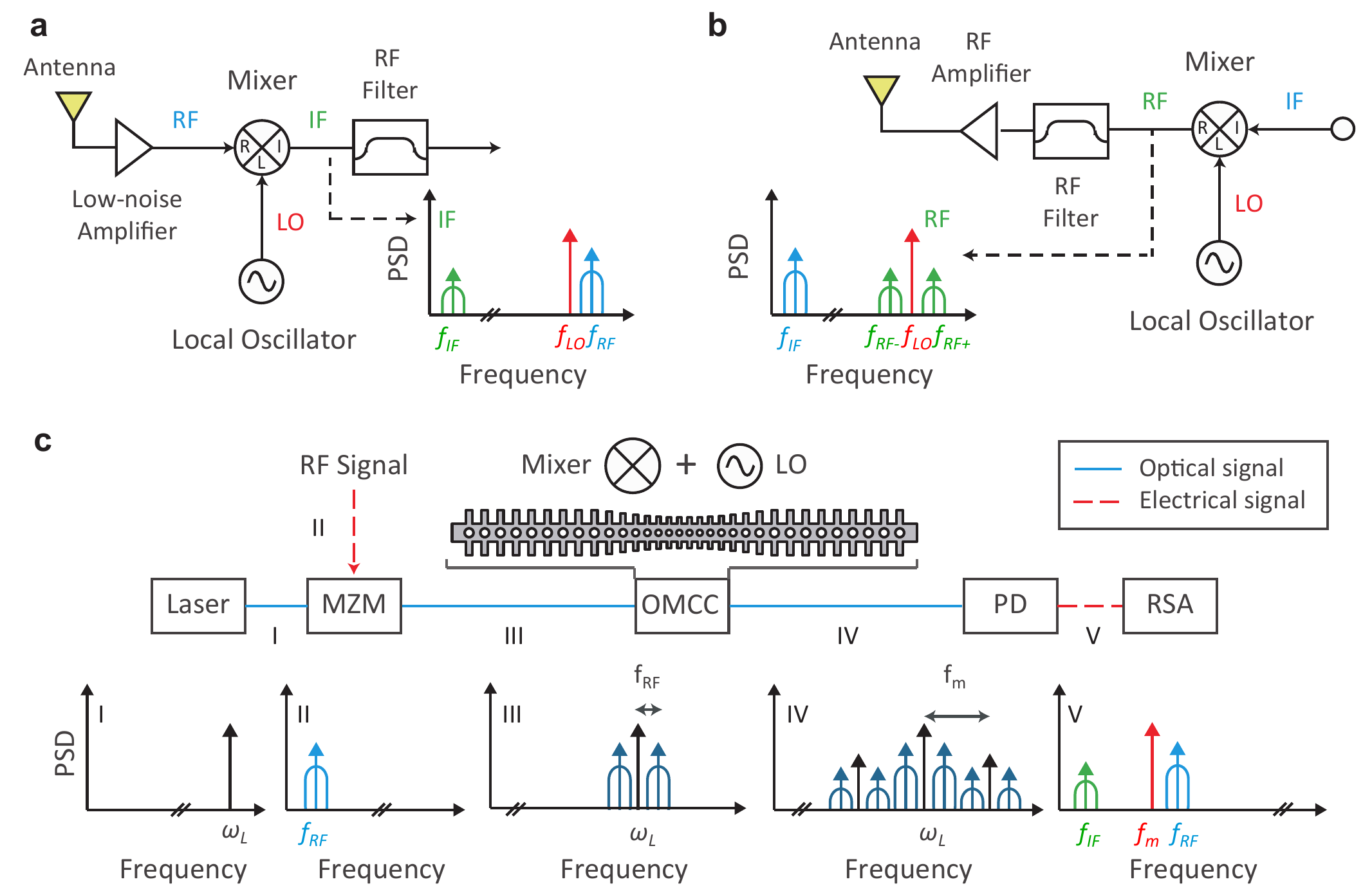}%
\end{center}
\caption{\textbf{RF frequency conversion process description. a} Frequency down-conversion and \textbf{b} up-conversion process carried out in the reception and transmission ends of a wireless system, respectively. Insets include the electrical spectra at the output of the mixer. In the up-conversion process, the RF signal is radiated, whilst in the down-conversion process the intermediate frequency (IF) signal --at a much lower frequency-- is processed internally. \textbf{c} Scheme of the photonic-based down-conversion process enabled by an OMCC, which acts simultaneously as an LO and a nonlinear mixer. %$f_{DS}$ accounts for the downconverted signal (DS) frequency.% 
The up-conversion process operates in the same way but interchanging the RF and IF signals. The LO frequency $f_{LO}$ is a harmonic of the mechanical frequency $m \times f_{m}$. MZM: Mach-Zehnder modulator; \hbox{OMCC: OM crystal cavity}; PD: Photodetector; RSA: RF Spectrum Analyser.}
\label{fig:scheme}
\end{figure*}

The use of OMCCs seems to be appropriate to satisfy all the previous requirements. In particular, no other approach enables photonics-based microwave processing in such reduced foot-prints, as can be seen in the scanning electron microscopy (SEM) image included in Fig. \ref{fig:motivation}b that shows an array of OMCCs built on a silicon chip. Recent experiments have unveiled the potential of OMCCs as ultra-compact all-optical microwave oscillators \cite{LUAN14-SR,GHOR19-ARX,MER20-NN, MER21-ARX}. Amongst other applications, microwave oscillators are widely used as LO to mix RF signal in wireless communication systems --as depicted in Fig. \ref{fig:scheme}a and b--. Interestingly, the inherent nonlinear response of OMCCs enables the self-sustained mixing of RF signals in the optical domain, being the LO generated intrinsically --at integer multiples of the mechanical resonance frequency-- as a result of the phonon lasing process in the cavity \cite{HOS10-IEEEJSTQE}. This means that, unlike other photonic approaches approaches, OMCCs enable frequency mixing without requiring an external LO (as depicted in detail in Fig. \ref{fig:scheme}c), which enormously simplifies the hardware requirements {(see Fig. \ref{fig:motivation}c)}.

Previously, OM-based frequency conversion was demonstrated through the down-conversion of 25 MHz tones modulating an optical signal using a silica microtoroid as OM cavity \cite{HOS08-IEEEPTL}. Frequency mixing at MHz frequencies was also demonstrated in a torsional OM resonator via dissipative coupling, which does not requires self-sustained oscillations \cite{HUA17-APL}. However, frequency conversion of real data streams modulated at GHz frequencies has not been addressed so far in OMCs to the best of our knowledge. In particular, frequency down- and up-conversion of digitally modulated signals employing orthogonal frequency division multiplexing (OFDM) modulation is of special interest. This is because OFDM is the underlying modulation in the most relevant communication systems, including 4G (e.g. LTE) and 5G cellular networks, as well as WiFi and WiMAX wireless access networks, and can be severely affected by the phase noise and frequency offset introduced in the conversion process \cite{LEH17-IEEE}. 

In this work, we demonstrate the all-optical frequency conversion of OFDM data streams in both down- and up-conversion processes using a silicon OMCC supporting a breathing-like mechanical mode at $f_m\approx$ 3.9 GHz that is coupled to an optical mode at telecom wavelengths. Operating the cavity in the phonon lasing regime at room temperature, we demonstrate experimentally that OFDM data signals following the IEEE 802.16e WiMAX standard can be down- and up-converted in frequency with a conversion efficiency better than -17 dB in both processes and without requiring an external LO. We also show that the strong non-linearity of the OMCC generates multiple harmonics of the mechanical resonance \cite{MER20-NN}, which can be used to increase the up-converted frequency above 10 GHz. This fact, added to the flexibility to tune the fundamental mechanical frequency by an appropriate design of the cavity \cite{OUD14-PRB}, would enable to adopt this approach in a wide variety of microwave systems and networks --including the MIMO systems displayed in Fig. \ref{fig:motivation}a-- enabled by all-optical processing in silicon chips.

\section{Results}

\subsection{Silicon OMCC cavity as a microwave oscillator}

\begin{figure*}[t]
\begin{center}
\includegraphics[width=0.7\linewidth]{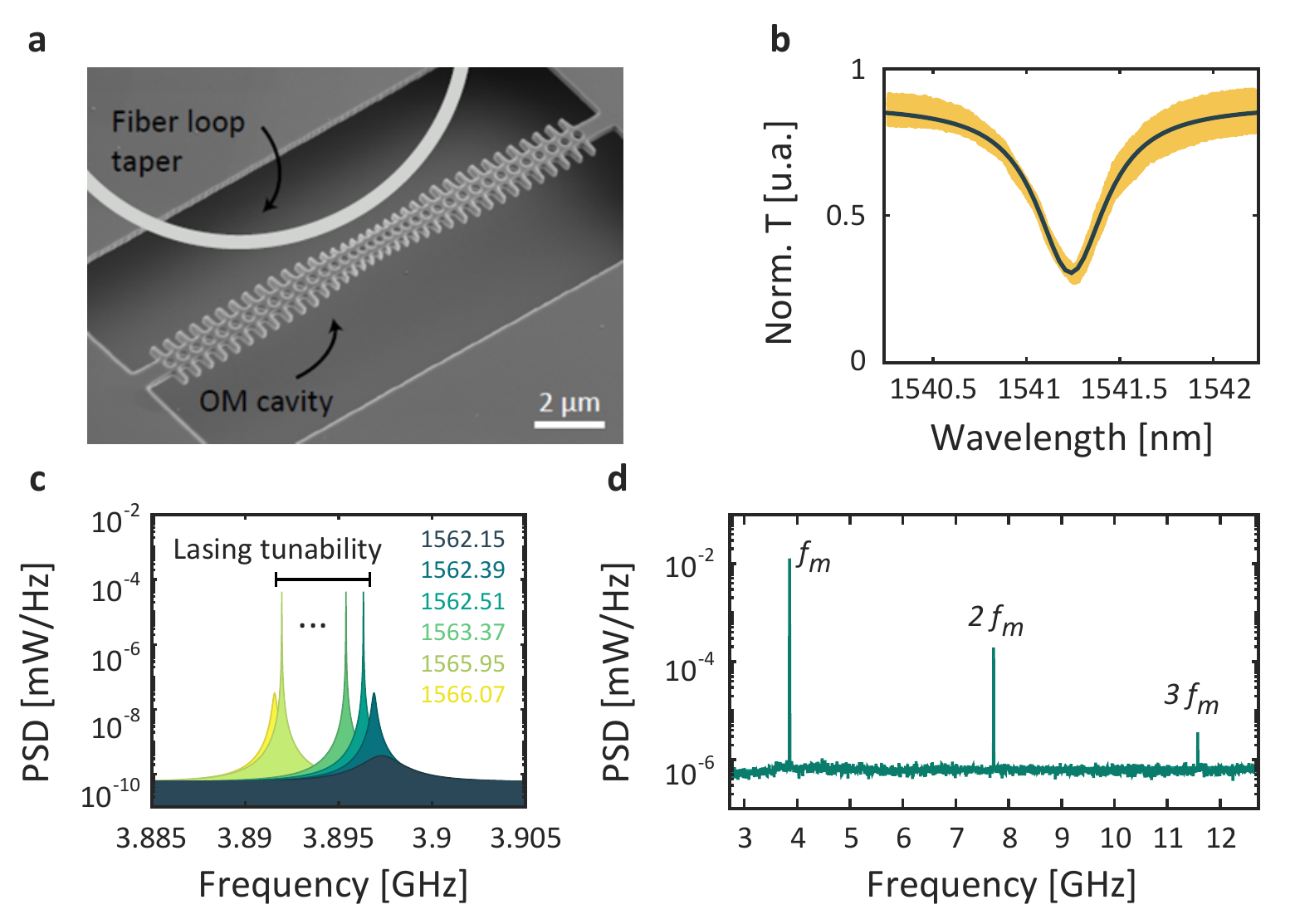}%
\end{center}
\caption{\textbf{Optical and mechanical properties of the silicon OM crystal cavity. a} SEM image of the tapered-fiber optical coupling to the OMCC under study. \textbf{b} Spectral optical response of the cavity in transmission. \textbf{c} Detected mechanical mode as a function of the driving wavelength, showing the transition from thermal noise transduction (darker colors) to the phonon lasing state (lighter colors) as well as the mechanical frequency tunability. \textbf{d} Detected frequency comb in the lasing state up to the third harmonic at a driving wavelength $\lambda=$1562.51 nm.}
\label{fig:omsystem}
\end{figure*}

Figure \ref{fig:omsystem}a shows a scanning electron microscopy (SEM) image of the OMCC used in our experiments. This cavity supports an optical mode at \hbox{$\lambda_{o}=$ 1541.2$\pm$0.3 nm} and a mechanical mode at $f_{m} \approx$ 3.9 GHz, both confined in the central region of the released nanobeam to ensure large OM interaction (the measured OM coupling rate was $g_{0}/2{\pi} \approx$ 660 kHz \cite{MER20-NN}). Noticeably, this OMCC is fabricated on a silicon-on-insulator wafer using standard fabrication processes \cite{NAV17-NCOMM} and it has a foot-print $\approx$ 10 $\mu$m$^{2}$, much smaller than other OM resonators such as silica microtoroids \cite{HOS10-IEEEJSTQE}. The optical response of the cavity at low input power ($P_{in} <$ 100 $\mu$W) when a tapered fiber is coupled to it via evanescent fields is depicted in Fig. \ref{fig:omsystem}b, showing a transmission dip that results in an optical Q factor $Q_{o} \approx$ 4 $ \times 10^{3}$ (see details about the experimental system in Supporting Information). 

\begin{figure}[t]
\center
\includegraphics[width=\linewidth]{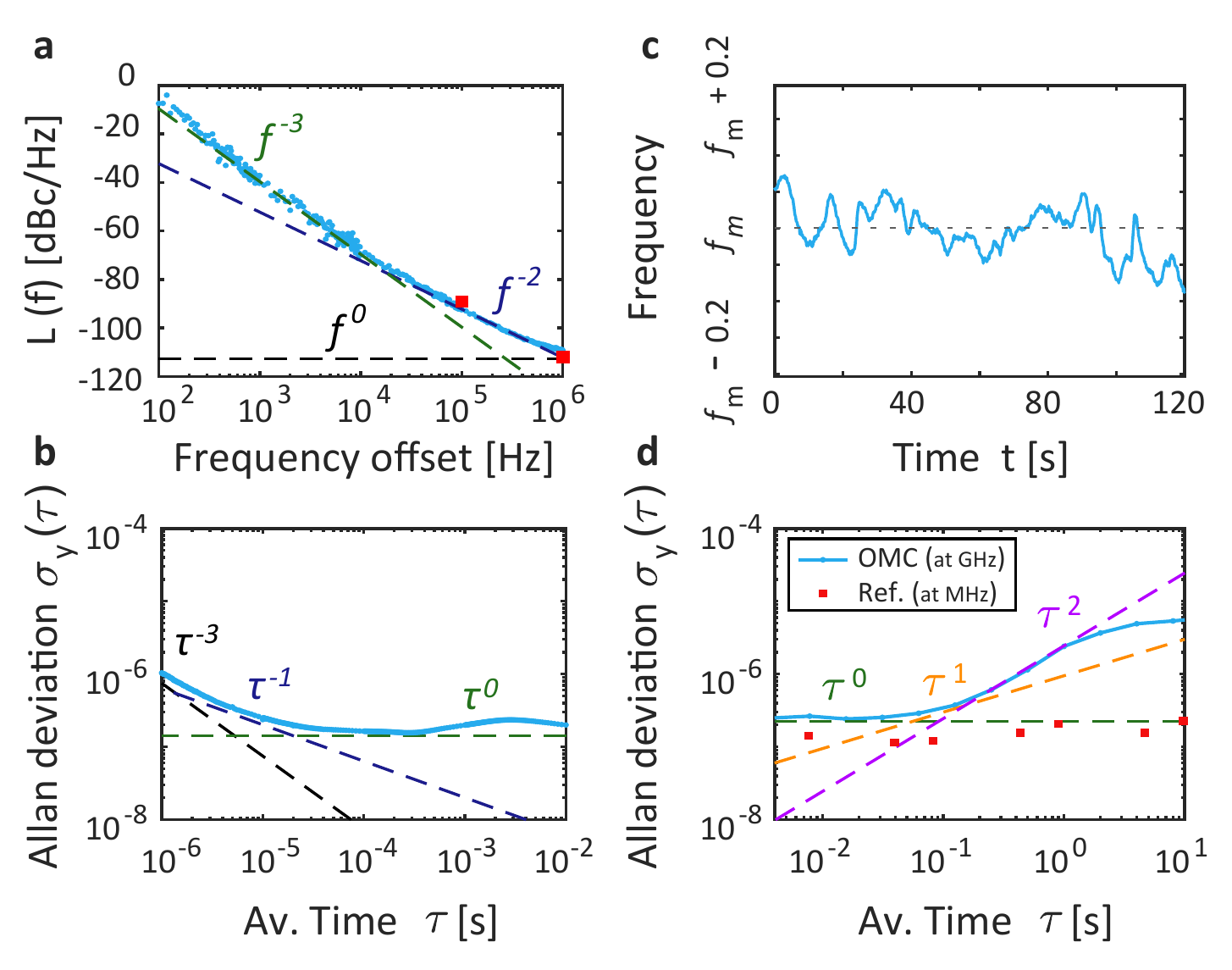}%
\caption{\textbf{OM microwave oscillator. a} Mean phase noise of the first harmonic (dark blue) and its uncertainty (light blue shaded area). As a reference, the phase noise values of WiMAX commercial equipment operating at 3.5 GHz frequency band of -89 dBc/Hz at 100 kHz offset and -112 dBc/Hz at 1 MHz offset are represented with red dots \cite{commercial}. \textbf{b} Short-term Allan deviation calculated from the phase noise measurement. \textbf{c} Evolution of the first harmonic frequency versus time within a span of 400 kHz. \textbf{d} Long-term Allan deviation calculated from the time evolution of the mechanical mode (blue) and non-stabilized nanomechanical beam at MHz frequencies (red dots)\cite{GAV13-NCOMM}. }
\label{fig:phasenoise}
\end{figure}

The mechanical motion of the cavity can be observed by photo-detecting the signal scattered from the cavity in the reflection path, as shown in Fig. \ref{fig:omsystem}c for different laser wavelengths with a driving power $P_{in}>$ 2.8 mW. As reported before \cite{NAV15-SR,MER20-NN}, scanning the wavelength across the blue side of the resonance allows to change the dynamics of the cavity going from the mere transduction of the thermal noise --darker curves in Fig. \ref{fig:omsystem}c-- to the phonon lasing regime --lighter curves in Fig. \ref{fig:omsystem}c--. Moreover, in this regime, a set of ultra-narrow tones at frequencies $m \times f_{m}$ is obtained --as shown in Fig. \ref{fig:omsystem}d-- which confirms the generation of an OM frequency comb in the cavity \cite{MER20-NN}. The total comb bandwidth is limited by the linewidth of the optical resonance. Remarkably, the phonon lasing state is maintained in a certain range of driving wavelengths --Fig. \ref{fig:omsystem}c--, although the observed tones are slightly shifted in frequency due to the optical spring effect \cite{ASP14-RMP}. This property could be eventually used to perform fine-tuning of the generated microwave tone.

Once the cavity is in the phonon lasing state, different parameters can be measured to characterize its performance as a photonics-based microwave oscillator. Figure \ref{fig:phasenoise}a depicts the mean phase noise of the first harmonic of the generated comb. The measured phase noise reaches values around \mbox{-100 dBc/Hz} at \mbox{100 kHz}, which is better than the requirement of commercial equipment employed to generate WiMAX signals (e.g. \mbox{-89 dBc/Hz} at \mbox{100 kHz} \cite{commercial}, marked as reference with red dots in Fig. \ref{fig:phasenoise}a). From this measurement, we can estimate the short-term Allan deviation $\sigma_{y}(\tau)$ \cite{RUB08-BOOK} (see Supporting Information), which is presented in \mbox{Fig. \ref{fig:phasenoise}b}. Here, different source noises can be identified such as the white phase noise ($L(f)\propto f^{0}$ or $\sigma_{y}(\tau)\propto \tau^{-3}$), the random walk of phase ($L(f)\propto f^{-2}$ or $\sigma_{y}(\tau)\propto \tau^{-1}$) and the flicker frequency noise ($L(f)\propto f^{-3}$ or $\sigma_{y}(\tau)\propto \tau^{0}$), in good agreement with observations in other OM oscillators operated in this regime \cite{HOS10-IEEEJSTQE, MER21-ARX}.

To estimate the long-term stability of the oscillator frequency -- an important parameter to evaluate the performance of the device in real applications --, we analyzed its time evolution, reported in Fig. \ref{fig:phasenoise}c. This measurement also permits the calculation of the Allan deviation for larger averaging times, as shown in Fig. \ref{fig:phasenoise}d. Here, we can observe that the flicker frequency noise contribution is at the same level as in the short-term Allan deviation in Fig. \ref{fig:phasenoise}b. Additionally, we can identify other noise sources such as a very small contribution of the random walk of frequency ($\sigma_{y}(\tau)\propto \tau^{1}$) and, finally, a linear frequency drift with $\sigma_{y}(\tau)\propto \tau^{2}$. Noticeably, these values are on par with the nanomechanical oscillator reported in \cite{GAV13-NCOMM} --represented in red dots in Fig. \ref{fig:phasenoise}d--, though our device operates at much higher frequencies. All these results lead us to conclude that our OM microwave oscillator could be used as all-optical processing element in real wireless systems. 

\subsection{OM frequency down-conversion}

%As stated above, the operating regime of our system is when the cavity is in the phonon lasing regime. In this situation, and under sufficient driving power, the output signals contains a series of harmonics around the laser wavelengths spaced by $f_{m}$, being the first harmonic characterized in Figs. \ref{fig:omsystem} and \ref{fig:phasenoise}. 
In our next experiment, we used the first harmonic of the mechanical resonance of the OM cavity operated in the phonon lasing regime to demonstrate OM frequency down-conversion. This process is commonly employed at the receiver side of a communication systems to decrease the frequency of the received data signal down to a band known as intermediate frequency (IF) where information can be easily processed (Fig. \ref{fig:scheme}a). To this end, the laser was modulated with an external microwave signal (the employed electrical power was \mbox{1 mW}) using an external Mach-Zehnder amplitude modulator. The external signal consisted of a pseudo-random sequence of data modulating a microwave signal (having a carrier frequency of \hbox{$f_{RF}$ = 3.9 GHz $> f_{m}$}) using orthogonal frequency division multiplexing (OFDM) with QPSK modulation per subcarrier considering bandwidths spanning from 3.5 MHz to 28 MHz, being fully compliant with the IEEE 802.16e standard used in wireless WiMAX networks \cite{IEEEwimax} (see details of the signal generation in Supporting Information). 

\begin{figure*}[t]
\begin{center}
\includegraphics[width=0.57\linewidth]{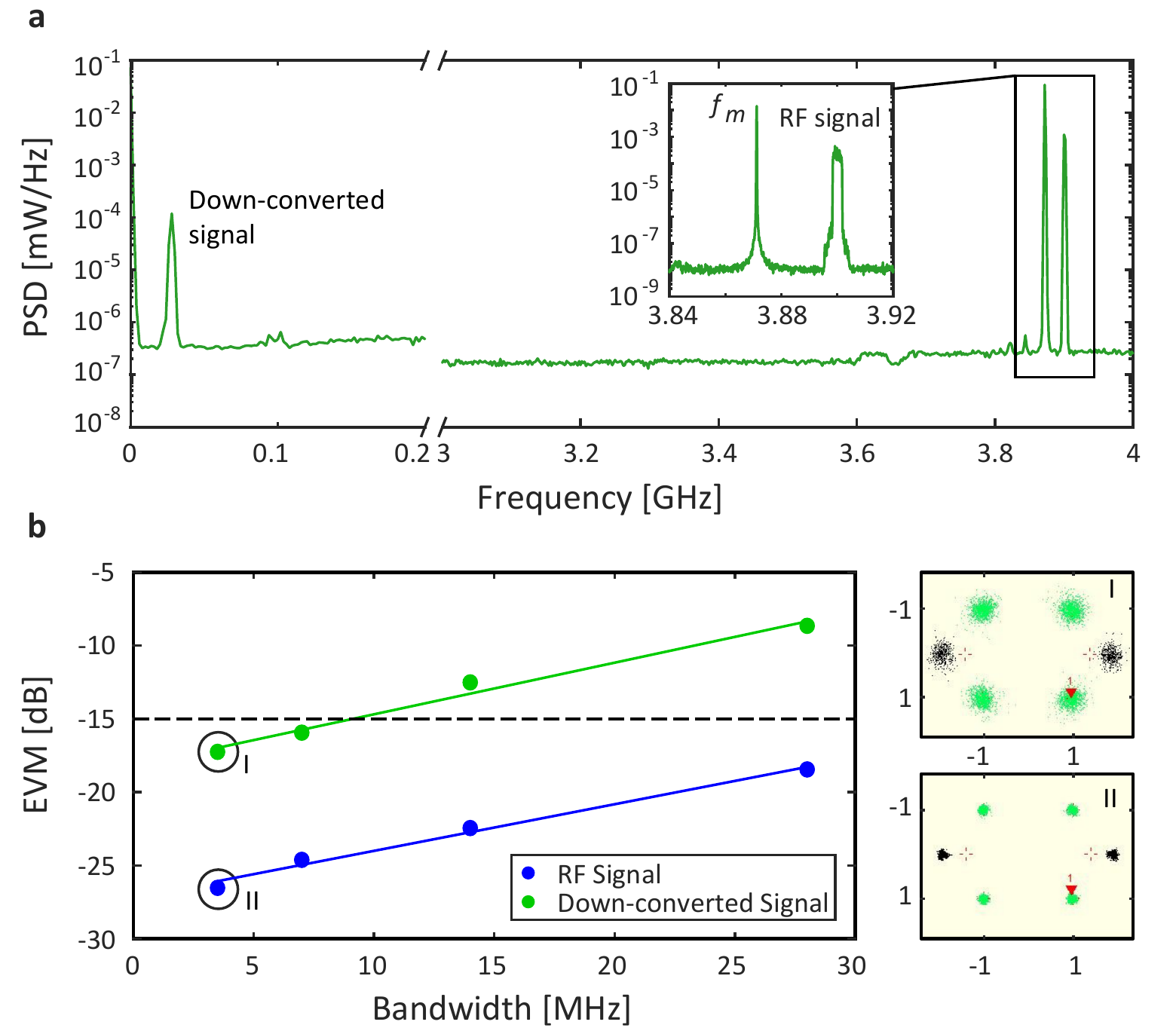}%
\end{center}
\caption{\textbf{OM frequency down-conversion. a} Detected electrical spectrum after down-conversion using the first harmonic of the mechanical resonance. (Inset) Close view of the RF signal and the mechanical mode in the lasing state. The resolution bandwidths are 1 MHz and 100 kHz in the main and inset panels, respectively. \textbf{b} Measured EVM vs. OFDM signal BW for the input RF signal (blue) and the down-converted IF signal (green). (Insets) Received QPSK constellations for the down-converted OFDM signal measured at $f_{IF}=f_{RF}- f_{m}\approx$ \hbox{50 MHz} (top panel) and the original RF signal at \hbox{$f_{RF}$ = 3.9 GHz} (bottom panel).}
\label{fig:downconversion}
\end{figure*}

The optical power reaching the cavity was 3.44 mW at $\lambda_{L}=$1562.51 nm, which was enough to drive the cavity into the phonon lasing regime. %(Laura, hay que dar todos los parámetros: potencia, frecuencias, lambda, e porofundidad de modulación; y super importante: estimar la eficiencia de down conversion: potencia del tono que se recibe vs. potencia del tono que se injecta). 
The measured RF spectrum is shown Fig. \ref{fig:downconversion}a, where a down-converted signal at a frequency $f_{IF}=f_{RF}- f_{m}\approx$ \hbox{50 MHz} is clearly observed. The efficiency of the down-conversion process is estimated to be around 3$\%$ ($\approx$ -15 dB) by comparing the power of the detected RF signal when operating out of the phonon lasing regime ($\approx$ 28 $\mu$W) and the power of the down-converted signal when operating in the phonon lasing regime ($\approx$ 0.85 $\mu$W). 

%If an input modulated tone is then injected into the system at a higher frequency ($f_{TONE} > f_{H1}$), this tone will be downconverted to a frequency corresponding to the different between the two involved tones ($f_{DOWN}=f_{TONE}-f_{H1}$). This process is presented in Fig. \ref{fig:downconversion}(a).

Figure \ref{fig:downconversion}b shows the evolution of the measured error vector magnitude (EVM) for the RF and the down-converted (IF) signals as a function of the OFDM signal bandwidth (BW). As expected, the EVM worsens as the signal bandwidth gets broader. However, a received EVM $<$ -15 dB, which is the threshold recommended at the received end in the standard \cite{IEEEwimax}, is obtained for signal bandwidths up to 7 MHz. In the inset panels, the received constellations for 3.5-MHz bandwidth OFDM signals for both the RF and the down-converted signals are presented. The green points in the constellation correspond to detected data symbols modulated in a certain subcarrier using QPSK modulation whilst the black points correspond to pilot signals using BPSK modulation. Remarkably, the recovery of QPSK data, as shown in the EVM measurements and the detected constellations, demonstrates that this technique is coherent since the phase is preserved in the frequency conversion process. In principle, improving the performance of the OM cavity would also allow satisfying the EVM threshold at higher bandwidths and higher modulation orders for the same BW as no phase distortion is observed in the received constellations. It should be noted that this OM oscillator does not have any feedback loop which could increase the stability and improve the signal to noise ratio. Additionally, a more efficient coupling of light into the cavity could be achieved by optimizing the fiber-cavity interface, which would also contribute to improving the signal to noise ratio and get better figures of the EVM.

\subsection{OM frequency up-conversion}

\begin{figure*}[t]
\begin{center}
\includegraphics[width=0.6\linewidth]{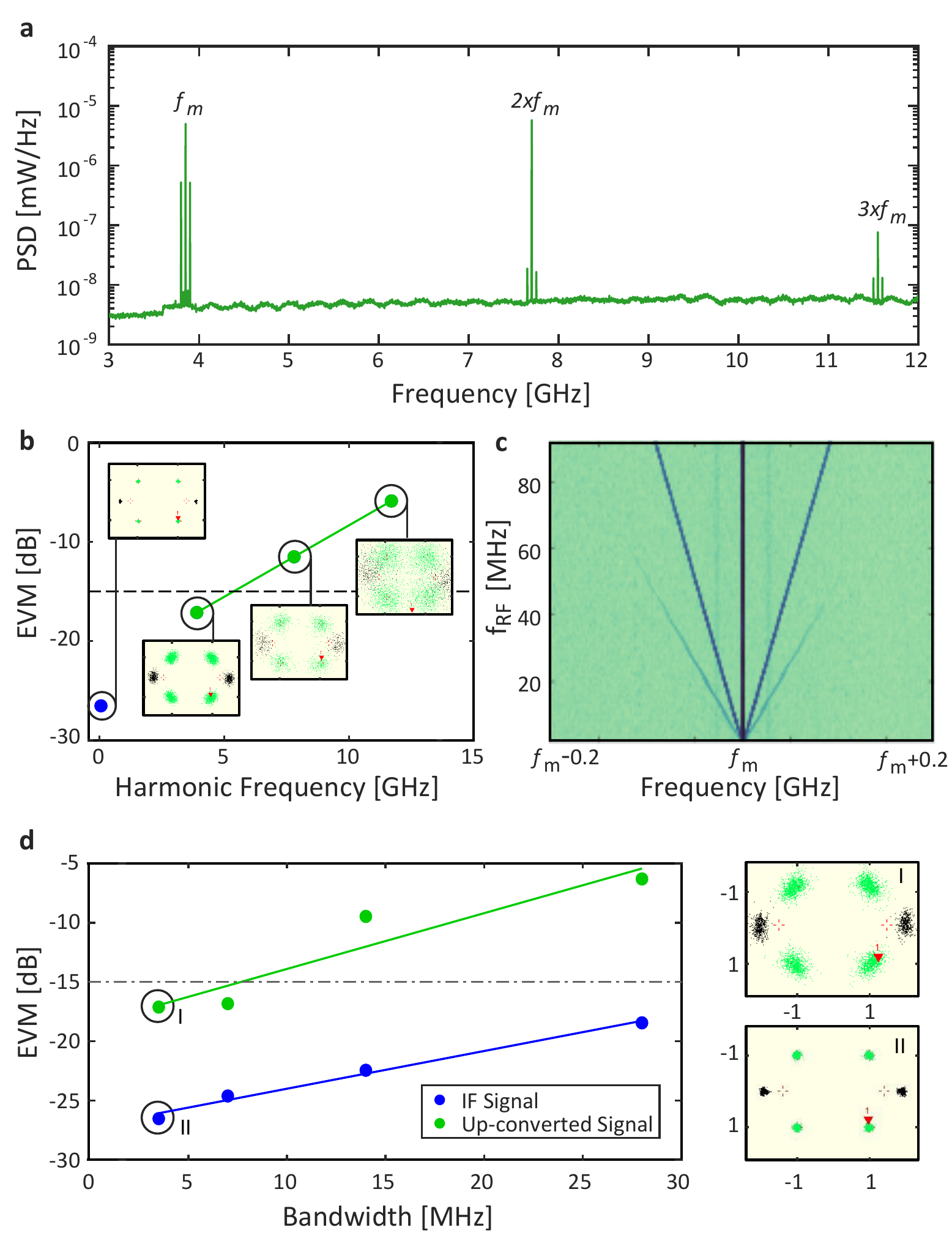}%
\end{center}
\caption{\textbf{OM frequency up-conversion. a} Measured electrical spectrum of the up-conversion process up to the third harmonic of the mechanical signal (resolution bandwidth = 1 MHz). \textbf{b} Measured EVM vs. harmonic frequency used in the up-conversion process. (Insets) Received QPSK constellations. \textbf{c} Evolution of the detected spectrum around the first harmonic as a function of the frequency of the modulation signal. \textbf{d} Measured EVM vs. OFDM signal bandwidth of the modulation signal (blue) and the 1st-harmonic up-converted signal (green). (Insets) Received constellations for the up-converted OFDM signal at $f_{RF}=f_{m}+f_{IF}$ and for the IF signal at \hbox{$f_{IF}=$ 50 MHz}. }
\label{fig:upconversion}
\end{figure*}

Since the OM cavity acts essentially as a nonlinear mixer with an intrinsic LO, it can also operate in reversely to the previous case and perform as an all-optical frequency up-converter. Frequency up-conversion is usually employed in communication systems and networks to increase the frequency of an incoming data stream up to a frequency band ($f_{RF}$) where it can be radiated (Fig. \ref{fig:scheme}b). To test the up-conversion process, the previous experiment is repeated employing actual OFDM data -- again compliant with the IEEE 802.16e WiMAX standard -- modulated now at a lower frequency (\mbox{$f_{IF}=$ 50 MHz}). The detected electrical spectrum in reflection is depicted in Fig. \ref{fig:upconversion}a, where lateral sidebands spaced $f_{IF}$ from the different harmonics of the mechanical resonance can be observed as a result of the mixing process taking place inside the cavity. The maximum observed frequency, corresponding to up-conversion using the third harmonic, is limited by the photodetector bandwidth (12 GHz). This result suggests that this approach, which may use also cavities supporting mechanical resonances at other frequencies, may cover a wide range of microwave frequencies and, as a result, be used in multiple wireless systems.

%Nota Alex: estos detalles los he puesto en la Conclusion. Al fin y al cabo, depende de la cavidad que se usae. Pero creo que queda claro que la técnica sirve para todo tipo de sistemas wireless.
%As depicted in Fig. \ref{fig:upconversion}(a), the first harmonic is measured at $f_{m}\approx$ 3.9 GHz, which lower sideband is suitable for upconversion in the regulated WiMAX band at 3.5-3.7 GHz and also for 3.4–3.8 GHz commercial 5G band \cite{5G}. The upper sideband of the first harmonic is suitable for 5.8-GHz WiMAX generation and for 5.2–5.8 GHz 5G NR band \cite{5G}. The second harmonic in Fig. \ref{fig:upconversion}(a) was measured experimentally at 2$\times f_{m}\approx$ 7.8 GHz, which lower sideband is suitable for recent 5G NR transmission in the mid-bands (sub-7) ranging from 5.9 to 7.1 GHz \cite{5G}. The third harmonic in Fig. \ref{fig:upconversion}(a) was measured at 3$\times f_{m} \approx$ 11.7 GHz, covering the X-band. 

The measured EVM for the up-converted OFDM signals using QPSK subcarrier modulation is presented in Fig. \ref{fig:upconversion}b for the original IF signal as well as different harmonics. Even though the -15 dB EVM recommendation at the receiver end is only met for the first harmonic (in this case in $f_{m}\approx$ 3.9 GHz), QPSK constellations can also be recovered for the second and third harmonics (see insets). As above, improving the signal to noise ratio of the up-converted signal should enable accomplishing the requirement. Still, being able to recover the phase of the \hbox{signal} is a signature of the coherence of the up-conversion process, even for higher-order harmonics. We also studied experimentally the tunability of the system. As shown in Fig. \ref{fig:upconversion}c, the up-conversion process works for different values of the input frequency, although the nonlinearity of the cavity also results in the formation of frequency terms at $n \times f_{IF}$. In principle, we should be able to up-convert IF signals with frequencies up to $f_{m}/2$ without overlapping with sidebands of higher-order harmonics. 

We also analyzed how the up-converted signal evolves as a function of the OFDM bandwidth for different signal bandwidths (from 3.5 MHz up to 28 MHz according to \cite{IEEEwimax}) for the first harmonic. As shown in Fig. \ref{fig:upconversion}d, the results look similar to the case of the down-conversion process: both 3.5 MHz and 7 MHz bandwidths satisfy the IEEE 802.16e standard EVM recommendation. This experimental result provides a proof-of-concept of the frequency conversion process for both down- and up-conversion applications with OFDM signals. In this case, the efficiency of the up-conversion process for the first harmonic is estimated to be around 2$\%$ ($\approx$ -17 dB) by comparing again the power of the detected RF signal presented above and the power of the up-converted signal when operating in the phonon lasing regime ($\approx$ 0.57 $\mu$W). Since in this case the up-converted power is distributed amongst different harmonics, it is reasonable to think that higher efficiencies in the first harmonic could be obtained by using cavities with larger Q factors so that higher-order harmonics are not generated.

\section{Conclusion}

We have demonstrated experimentally that OMCCs can be successfully used as nonlinear elements capable to perform down- and up-conversion of digitally modulated signals employing OFDM modulation at frequencies around 4 GHz with MHz-bandwidths. This functionality would enormously simplify the optoelectronic hardware in wireless systems and networks since the OMCC acts simultaneously as a mixing element and as an LO. Remarkably, the whole process is coherent, which is demonstrated through the successful recovery of OFDM signals following the IEEE 802.16e WiMAX wireless standard after down- and up-conversion. This is of special importance for massive MIMO transmissions with a large number of AEs, which may be eventually addressed by using arrays of OMCCs built on a same chip. With a measured phase noise of around -100 dBc/Hz at 100 kHz, the proposed method meets the requirements of commercial wireless equipment. A frequency conversion efficiency greater than -17 dB was obtained in both down- and up-conversion processes.

The device, which is built on a silicon chip using standard microfabrication tools, has a reduced foot-print (\hbox{$\approx$ 10 $\mu$m$^{2}$}) in comparison with other OM resonators. It can operate at the fundamental mechanical frequency as well as at integer multiples of it, so wireless systems operating at different spectral bands through the whole radio-frequency spectrum could be benefited from this approach. For instance, using the first harmonic, the up-conversion frequencies would span from $f_{m}-f_{m}/2$ to $f_{m}+f_{m}/2$. Therefore, our cavity would cover the regulated WiMAX bands at 3.5-3.7 GHz and 5.8 GHz, and would also be suitable for the 3.4-3.8 GHz and \hbox{5.2-5.8 GHz} commercial 5G new-radio (NR) bands \cite{5G}. Up-converted frequencies can be extended using higher-order harmonics: the second harmonic, measured at 7.8 GHz, would be suitable for 5G NR transmission in the mid-bands, as recently demonstrated (sub-7)\cite{5G} and the third harmonic measured at 11.7 GHz reaches the technologically relevant X-band. Besides this, other OM cavities could also be used for frequency mixing at other frequencies. Indeed, OM crystals exhibit a tremendous flexibility in the design of the mechanical frequency, which can go from tens of MHz (using the flexural motion of the released silicon beam as in \cite{NAV15-SR}) up to around 10 GHz using more complex two-dimensional OM crystals \cite{REN20-ARX}. Besides its use in photonics-enabled MIMO-based wireless communications systems, our work shows that OM cavities integrated in silicon chips hold the potential to produce a dramatic reduction in size, weight and power consumption of microwave photonics devices, which is absolutely crucial in aircraft and satellite communication systems. 

Still, some issues need to be addressed in the way towards practical applications. The cavity-fiber interface needs to be improved mainly in terms of robustness to avoid fluctuations and variability arising from the current coupling technique. The use of a waveguide with an inverted taper should ensure easy and efficient coupling to an external lensed fiber that could be positioned in a V-groove to increase stability \cite{MEE14-PRA}. The need for a released silicon beam makes the structure inherently more fragile that standard silicon photonic devices in which the silicon core rests over a silica substrate. Interestingly, recent approaches suggest the feasibility of building GHz-frequencies OM circuits in an unreleased platform compatible with silicon technology \cite{MER21-PRA}. Finally, low-cost production, a main feature of silicon photonics, would require fabrication in large volumes using CMOS technology. Interestingly, recent experiments show that high-quality OMCCs can be fabricated via optical lithography \cite{BEN17-SR}, as in CMOS foundries, in contrast to the electron-beam lithography process widely used to fabricate small series of devices.

\medskip
\textbf{Supporting Information} \par %Please delete the Suppporting Information statement if it is not applicable. Please supply Supporting Information in another file. Supporting information should not be provided in .tex format
Supporting Information is available from the author.

% Acknowledgements
\medskip
\textbf{Acknowledgements} \par %delete if not applicable))
Funding: The authors acknowledge funding from the H2020 Future and Emerging Technologies program (projects PHENOMEN 713450 and SIOMO 945915); the Spanish State Research Agency (PGC2018-094490-BC21, PID2019-106163RJ-I00/AEI/10.13039/501100011033 MULTICORE+ and MCIU/AEI/FEDER UE RTI2018-101296-B-I00 MULTI-BEAM5G); Generalitat Valenciana (PPC/2021/042, BEST/2020/178, PROMETEO/2019/123 and IDIFEDER/2018/033); and the UPV Programa de Ayudas de Investigaci\'on y Desarrollo (PAID-01-16).

% References
\medskip

\end{document}